\documentclass[prd, preprint]{revtex4-1}
\usepackage{graphicx}
\usepackage{amsmath,amssymb}

\begin{document}

\title{On Teleparallel Quantum Gravity in Schwarzschild Space-Time}

\author{S. C. Ulhoa}\email[]{sc.ulhoa@gmail.com}
\affiliation{Instituto de F\'{i}sica,
Universidade de Bras\'{i}lia, 70910-900, Bras\'{i}lia, DF,
Brazil.\\Faculdade Gama, Universidade de Bras\'{i}lia, Setor Leste
(Gama), 72444-240, Bras\'{i}lia-DF, Brazil.}
\author{R.G.G. Amorim}\email[]{ronniamorim@gmail.com}
\affiliation{Instituto de F\'{i}sica,
Universidade de Bras\'{i}lia, 70910-900, Bras\'{i}lia, DF,
Brazil.\\Faculdade Gama, Universidade de Bras\'{i}lia, Setor Leste
(Gama), 72444-240, Bras\'{i}lia-DF, Brazil.}

\begin{abstract}
In this article we present the quantization process for
Schwarzschild space-time in the context of Teleparallel gravity. In
order to achieve such a goal we use the Weyl formalism that
establishes a well defined correspondence between classical quantities which are realized by functions and quantum ones which are realized by operators. In the process of
quantization we introduce a fundamental constant that is used to
construct what we call the quantum of matter by the imposition of
periodic conditions over the eigenfunction.
\end{abstract}

\maketitle

\section{Introduction}

The dynamical behavior of physical systems can be realized essentially by two descriptions of reality, the classical approach to which the physical world evolves deterministicly and the quantum one which describes nature by means the concept of probabilities.
The quantum description is usually obtained from classical
description through appropriate processes, the so called quantization procedures~\cite{berezin1975,2005RvMaP..17..391A,Giulini:2003mn}. The first ideas about quantization emerged in 1925, with
Heisenberg who proposed a description of quantum
mechanics based solely in terms of observable quantities~\cite{French:1985:NBC}. Thus Heisenberg used an amplitude multiplication rule that
later Born has identified to a matrix calculation~\cite{Hund}. In such a formalism scope, important results have
been yielded, among them we point out  the quantization of the  harmonic oscillator solution
and the achievement of a
commutation relation between position and its conjugated momentum. Born
included Jordan in such a discussion and together they have
generalized what was known so far for systems with arbitrary
degrees of freedom, they have introduced the canonical
transformations for this context as well~\cite{Hund}. In 1926, Pauli gave his contribution to the development of the quantization procedure by showing how to obtain the hydrogen
spectrum from this formalism~\cite{Hendry}. Dirac, independently, was able to
establish the connection between classical and quantum mechanics,
relying on the Hamilton-Jacobi formulation of classical
mechanics and using an algebraic formalism\cite{Hendry, citeulike:11602931}. Born and
Wiener have focused on  matrix approach which has led to the representation of the Hamiltonian (until then a classical function) in terms of operators, in this
sense, arose the first quantization procedure \cite{French:1985:NBC,
citeulike:11602931}.

Since then, the process of quantizing a physical system has became a controversial subject and several methods have been proposed, among them stand out the
canonical quantization, path integral quantization and Weyl
quantization which will be focused in this paper. The first two methods are based on Dirac rules and Feynman generating functional respectively~\cite{Ali}.
Both have some problems such as the non-invariance under canonical transformations and they seem to be not extendable to
non-euclidian phase-spaces. Particularly the canonical quantization method
leads to difficulties in the understanding  of the quantum-classical
limit. On the other hand, the Weyl quantization procedure, developed in 1927 \cite{weyl},  is a more robust approach,
in such a method there is a well defined mathematical operation with a clear
correspondence rule between classical functions and quantum
operators. Even though these methods are largely used to quantize a classical field, such processes are far from being unanimously accepted, for instance the arising operators order is very controversial~\cite{Shewell, Ali}.

We would like to implement the Weyl quantization procedure to
construct a quantum theory of gravitation. So far every attempt to
address the problem of quantum gravity is based on General
Relativity which is the most receptive theory of gravitation in the
scientific community. Such an approach revealed to be ill defined,
for instance we point out the problem of time in loop quantum
gravity~\cite{Anderson:2010xm} and the non-renormalization
problem~\cite{Shiekh:1996fd,Shomer:2007vq}. In our opinion those problems arise from the fact that General Relativity is not a self-consistent theory since it presents some difficulties that has not been overcome over the years such as the problem of gravitational energy~\cite{Landau1980Classical,Komar,Maluf:1995re}. Therefore we shall work with an alternative
theory of gravitation, the so called Teleparallelism Equivalent to
General Relativity. The reason for such a choice is very simple
indeed: Teleparallel gravity allows the existence of a gravitational
energy-momentum vector. Such a feature is not present in General
Relativity, although both theories are equivalent when it comes to
the dynamics of the gravitational field. The geometry in which
Teleparallel gravity is constructed is richer than the Riemannian
geometry, this yields a wider point of view for Teleparallelism in
the analysis of what is going on in the space-time, mainly in the
definition of conserved quantities. Teleparallel gravity has been
developed and tested over the years in what concerns its classical
features\cite{Hehl,Hehl2,Shirafuji,Maluf:2006gu,Maluf:2006gu,Maluf:2008ug,Maluf:2008yy,ANDP:ANDP201000168} and in our opinion it seems to be a
plausible theory of gravitation. However there are few attempts to quantize this
theory, for instance we refer the set of papers \cite{0264-9381-30-19-195003,Okolow:2013ifa,Okolow:2013dua,Okolow:2013lba} which were developed as an application of Dirac's method to TEGR. Hence we intent to give our contribution in this process by analyzing the quantum version of Schwarzschild's solution of field
equations that arises from the identification ${\cal H}=e\,t^{(0)0}$ in the Weyl's prescription.

The paper is organized as follows. In the section \ref{tf}, the Weyl
quantization procedure is detailed and some basic ideas of
Teleparallel gravity are presented. In section \ref{qg}, we develop
our version of a quantum theory of gravitation for Schwarzschild's
solution. Thus we introduce a new fundamental constant, necessary
for the quantization procedure, which leads to the definition of a
quantum of matter. Finally we present our concluding remarks.

\bigskip
Notation: space-time indices $\mu, \nu, ...$ and SO(3,1) indices $a,
b, ...$ run from 0 to 3. Time and space indices are indicated
according to $\mu=0,i,\;\;a=(0),(i)$. The tetrad field is denoted by
$e^a\,_\mu$ and the determinant of the tetrad field is represented
by $e=\det(e^a\,_\mu)$. The tetrad field is related to the metric by $e^a\,_{\mu}e_{a\nu}=g_{\mu\nu}$. In addition we adopt units where
$G=c=1$, unless otherwise stated.\par
\bigskip

\section{Theoretical Framework}\label{tf}

\subsection{Weyl Quantization}

In this subsection, we present a quantization procedure called Weyl quantization. We would like to remark that Weyl quantization, in opposite to canonical procedure, is a well defined mathematical framework  and can be extended to study of non-euclidian phase spaces. Furthermore, using Weyl procedure we can observe easily the correspondence principle. In this sense, we consider a classical system described by $n$ variables which will be denoted by $z_1$, $z_2$,$...$, $z_n$, those variables would be quantized by the prescription
\begin{equation}\nonumber
(z_1,z_2,...,z_n)\rightarrow (\widehat{z}_1,\widehat{z}_2,...,\widehat{z}_n).
\end{equation}
When the classical variables  $z_1$, $z_2$,$...$, $z_n$  are quantized by the above rule, the functions $f$ defined on those variables are immediately quantized. This quantization of the functions $f$ occurs by Weyl's map, $\mathcal{W}: f\rightarrow \widehat{f}=\mathcal{W}[f]$, which is given by
\begin{equation}\label{w1}
 \mathcal{W}[f](z_1, z_2,..., z_n):= \frac{1}{(2\pi)^n}\int d^n k d^n z f(z_1, z_2,..., z_n)\exp{\left(i\sum_{l=1}^{n} k_l(\widehat{z}_l - z_l)\right)}.
\end{equation}
This quantum-classical correspondence is called Weyl quantization. Under the formal mathematical viewpoint, the Weyl method is used to formulate the Groenewold-Moyal quantum mechanics~\cite{2004EJPh...25..525H}.
The kernel of this transformation is given by
\begin{equation}\nonumber
\Delta(\widehat{z}, z)=\frac{1}{(2\pi)^n}\int d^n k \exp{\left(i\sum_{l=1}^{n} k_l(\widehat{z}_l - z_l)\right)}.
\end{equation}
In this way, the Weyl map is written by
\begin{equation}\nonumber
\mathcal{W}[f](z_1, z_2,..., z_n):= \int  d^n z \Delta(\widehat{z},z) f(z_1, z_2,..., z_n).
\end{equation}
Formally the set of operators $\widehat{z}_i$ form a non-commutative space~\cite{weyl}. The construction of this non-commutative space is given by the replacement of local coordinates $z_i$ by the Hermitian operators $\widehat{z}_i$, which leads to the following commutation relation
\begin{equation}\nonumber
[\widehat{z}_i, \widehat{z}_j]=i\alpha_{ij},\label{nc}
\end{equation}
where $\widehat{z}_i$ are operators of a noncommutative algebra and $\alpha_{ij}$ is an anti-symmetric tensor.
Thus, the product of two operators in non-commutative space is given by $ \mathcal{W}[f(z)]\mathcal{W}[g(z)]=\mathcal{W}[f(z)\star f(z)]$ where the Moyal (or star) product is defined by
\begin{equation}\nonumber
f(z)\star g(z)= f(z)\exp\left[\frac{i}{2}\alpha^{ij}\overleftarrow{\partial}_i\overrightarrow{\partial}_j\right]g(z).
\end{equation}
As an example let us consider the function $f(z_1,z_2)=z_1^2+2z_1z_2+z_2^2$. Applying the Weyl procedure in this function we obtain
\begin{equation}\nonumber
\widehat{f}(\widehat{z}_1,\widehat{z}_2)=\widehat{z}_1^2+\widehat{z}_1\widehat{z}_2+\widehat{z}_2\widehat{z}_1+\widehat{z}_2^2.
\end{equation}
We realized that the Weyl quantization eliminates the ambiguity in variables ordering present in canonical procedure. Another advantage in Weyl procedure is its  use in the quantization of non-polynomial functions~\cite{Mukunda:2005us}.

\subsection{Teleparallel
Equivalent to General Gelativity (TEGR)}\label{tel} \noindent

In this subsection we briefly recall the ideas concerning Teleparallel gravity which can be tracked back to the 1930's when Einstein made an attempt to unify gravitation and electromagnetism~\cite{einstein}. In this theory the dynamics of the field relies on the tetrads, $e^{a}\,_{\mu}$, rather than on the metric tensor, $g_{\mu\nu}$. It can be formally described by means a Weitzenb\"ock geometry ~\cite{Cartan}, in which the Cartan connection
$\Gamma_{\mu\lambda\nu}=e^{a}\,_{\mu}\partial_{\lambda}e_{a\nu}$, plays a central role. Thus the torsion associated to such a connection is given by

\begin{equation}
T^{\mu}\,_{\lambda\nu}=e_a\,^{\mu}\left(\partial_{\lambda} e^{a}\,_{\nu}-\partial_{\nu}
e^{a}\,_{\lambda}\right)\,, \label{3}
\end{equation}
or simply $T^{\mu}\,_{\lambda\nu}=e_a\,^{\mu}T^{a}\,_{\lambda\nu}$, where $T^{a}\,_{\lambda\nu}=\partial_{\lambda} e^{a}\,_{\nu}-\partial_{\nu}e^{a}\,_{\lambda}\,.$

We intend to show the equivalence between General Relativity and Teleparallel gravity by showing the equivalence between the geometrical framework of both theories. Firstly we note that the Christoffel symbols
(${}^0\Gamma_{\mu \lambda\nu}$) yield a vanishing torsion tensor due to its symmetric features. The Cartan connection and the Christoffel symbols are related by the following mathematical identity

\begin{equation}
\Gamma_{\mu \lambda\nu}= {}^0\Gamma_{\mu \lambda\nu}+ K_{\mu
\lambda\nu}\,, \label{2}
\end{equation}
where

\begin{eqnarray}
K_{\mu\lambda\nu}&=&\frac{1}{2}(T_{\lambda\mu\nu}+T_{\nu\lambda\mu}+T_{\mu\lambda\nu})\label{3.5}
\end{eqnarray}
is the contortion tensor. In the same way the Cartan connection yields a vanishing scalar curvature. Thus in the Weitzenb\"ock geometry there is a vanishing curvature while in the Riemann geometry there is a vanishing torsion. Both geometries are related by expression (\ref{2}), from which it is possible to obtain the relation

\begin{equation}
eR(e)\equiv -e({1\over 4}T^{abc}T_{abc}+{1\over
2}T^{abc}T_{bac}-T^aT_a)+2\partial_\mu(eT^\mu)\,,\label{5}
\end{equation}
where $e$ is the determinant of the tetrad field, $T_a=T^b\,_{ba}$ and $R(e)$ is the
scalar curvature constructed out in terms of such a field. Therefore we choose the Lagrangian density, in the realm of Teleparallel gravity, as

\begin{equation}
\mathfrak{L}= -k e({1\over 4}T^{abc}T_{abc}+{1\over
2}T^{abc}T_{bac}- T^aT_a) -\mathfrak{L}_M \,,
\end{equation}
where $k=1/16\pi$ and $\mathfrak{L}_M$ stands for the Lagrangian density
for the matter fields. It worths to mention that the total divergence had been dropped out in the construction of the Lagrangian density, since it do not contribute to the field equations. It also should be noted, from (\ref{5}), that the geometrical part of this Lagrangian density is exactly the Hilbert-Einstein Lagrangian density. Hence both theories share the same dynamical properties. However in Teleparallel gravity it is possible to define a gravitational energy-momentum tensor. Let us rewritten the Lagrangian density as

\begin{equation}
\mathfrak{L}\equiv -ke\Sigma^{abc}T_{abc} -\mathfrak{L}_M\,,
\label{5.1}
\end{equation}
where

\begin{equation}
\Sigma^{abc}={1\over 4} (T^{abc}+T^{bac}-T^{cab}) +{1\over 2}(
\eta^{ac}T^b-\eta^{ab}T^c)\,. \label{6}
\end{equation}
Then the field equations can be derived from (\ref{5.1}) using a
variational derivative with respect to $e^{a\mu}$, they read

\begin{equation}
e_{a\lambda}e_{b\mu}\partial_\nu (e\Sigma^{b\lambda \nu} )- e
(\Sigma^{b\nu}\,_aT_{b\nu\mu}- {1\over 4}e_{a\mu}T_{bcd}\Sigma^{bcd}
)={1\over {4k}}eT_{a\mu}\,, \label{7}
\end{equation}
where $\delta \mathfrak{L}_M / \delta e^{a\mu}=eT_{a\mu}$. Those equations may be rewritten as

\begin{equation}
\partial_\nu(e\Sigma^{a\lambda\nu})={1\over {4k}}
e\, e^a\,_\mu( t^{\lambda \mu} + T^{\lambda \mu})\;, \label{8}
\end{equation}
where $T^{\lambda\mu}=e_a\,^{\lambda}T^{a\mu}$ and

\begin{equation}
t^{\lambda \mu}=k(4\Sigma^{bc\lambda}T_{bc}\,^\mu- g^{\lambda
\mu}\Sigma^{bcd}T_{bcd})\,. \label{9}
\end{equation}
In view of the antisymmetry property
$\Sigma^{a\mu\nu}=-\Sigma^{a\nu\mu}$, it follows that

\begin{equation}
\partial_\lambda
\left[e\, e^a\,_\mu( t^{\lambda \mu} + T^{\lambda \mu})\right]=0\,,
\label{10}
\end{equation}
which is local balance equation. Therefore such equation leads to the following continuity equation,

\begin{equation}
{d\over {dt}} \int_V d^3x\,e\,e^a\,_\mu (t^{0\mu} +T^{0\mu})
=-\oint_S dS_j\, \left[e\,e^a\,_\mu (t^{j\mu} +T^{j\mu})\right]\,.
\label{11}
\end{equation}
Thus we identify $t^{\lambda\mu}$ as the gravitational
energy-momentum tensor~\cite{PhysRevLett.84.4533,maluf2}.

Then, as usual, the total energy-momentum vector is defined
by~\cite{Maluf:2002zc}

\begin{equation}
P^a=\int_V d^3x\,e\,e^a\,_\mu (t^{0\mu} +T^{0\mu})\,, \label{12}
\end{equation}
where $V$ is a volume of the three-dimensional space. We point out that the energy-momentum vector is invariant under coordinates transformations and it is sensible to frame transformations as it should be expected.

\section{Quantum Gravity}\label{qg}

In this section we address the problem of quantization of gravity in the framework of Teleparallelism Equivalent to General Relativity. Then we start with a stationary space-time

\begin{equation}
ds^2=g_{00}dt^2+g_{11}dr^2+g_{22}d\theta^2+g_{33}d\phi^2\,,\label{metrica}
\end{equation}
we bring attention to the fact that this line element is written in spherical coordinates and the metric tensor components are functions of $r$ and $\phi$, solely. In addition we point out that $g_{00}<0$, thus the metric tensor has the proper limit as Minkowski space-time.

There are an infinite number of possible tetrads satisfying the relation $g_{\mu\nu}=e^a\,_\mu e_{a\nu}$ for (\ref{metrica}). To fix it we interpret the tetrad field as being a reference frame adapted to a observer in space-time. Thus we choose

\begin{equation}
e^a\,_{\mu}=\left(
  \begin{array}{cccc}
    \sqrt{-g_{00}}&0&0&0 \\
    0&\sqrt{g_{11}}
\,\sin\theta \cos\phi& \sqrt{g_{22}} \cos\theta \cos\phi & -\sqrt{g_{33}}\sin\phi \\
    0& \sqrt{g_{11}}\, \sin\theta \sin\phi&
\sqrt{g_{22}} \cos\theta \sin\phi &  \sqrt{g_{33}} \cos\phi \\
    0&
\sqrt{g_{11}}\, \cos\theta & -\sqrt{g_{22}}\sin\theta&0 \\
  \end{array}
\right)\,,\label{tetrada}
\end{equation}
which is adapted to a stationary observer~\cite{Maluf:2007qq}. In order to obtain the gravitational energy, firstly we need to obtain the $\Sigma^{(0)0i}$ components, they read

\begin{eqnarray}
4e\Sigma^{(0)01}&=&2\left(\sqrt{g_{33}}+\sqrt{g_{22}}\sin\theta\right)-\frac{1}{\sqrt{g_{11}}}\left[\sqrt{\frac{g_{33}}{g_{22}}}\left(\frac{\partial g_{22}}{\partial r}\right)+\sqrt{\frac{g_{22}}{g_{33}}}\left(\frac{\partial g_{33}}{\partial r}\right)\right]\nonumber\,,\\
4e\Sigma^{(0)02}&=&2\sqrt{g_{11}}\cos\theta-\frac{1}{\sqrt{g_{22}}}\left[\sqrt{\frac{g_{11}}{g_{33}}}\left(\frac{\partial g_{33}}{\partial \theta}\right)+\sqrt{\frac{g_{33}}{g_{11}}}\left(\frac{\partial g_{11}}{\partial \theta}\right)\right]\,,\nonumber\\
e\Sigma^{(0)03}&=&0\,.
\end{eqnarray}
We restrict our attention to Schwarzschild space-time to which $g_{00}=\left(1-\frac{2M}{r}\right)=g_{11}^{-1}$, where $M$ is the black hole mass. Thus the only non-vanishing $\Sigma^{(0)0i}$ component reads

$$4e\Sigma^{(0)01}=4r\sin\theta\left[1-\left(1-\frac{2M}{r}\right)^{1/2}\right]\,.$$

We recall that $E\equiv P^{(0)}$, then we have

$$E=4k\int d^3x\partial_i\left(e\Sigma^{(0)01}\right)\,,$$
which can be represented by $E=\int d^3x {\cal H}$. Hence ${\cal H}=4k\partial_i\left(e\Sigma^{(0)01}\right)$ which for Schwarzschild space-time yields

\begin{equation}
{\cal H}=4k\sin\theta\left[1-\frac{\left(1-\frac{M}{r}\right)}{\left(1-\frac{2M}{r}\right)^{1/2}}\right]\,.
\end{equation}
This is the classical (non-quantum) gravitational Hamiltonian density, it should be noted that it is a tensorial density and as consequence it transforms accordingly under coordinate transformations.

The procedure to quantize this field is formally given by the Weyl's prescription which is the following
$\theta\rightarrow\widehat{\theta}$ and $r\rightarrow\widehat{r}$, where $\widehat{\theta}=i\alpha\frac{\partial}{\partial r}$ and
$\widehat{r}=r$. Here $\alpha$ is a constant with dimension of distance. Thus the commutator between such operators is

$$[\widehat{\theta},\widehat{r}]=i\alpha\,,$$ as defined by relation
(\ref{nc}). As a consequence ${\cal H}\rightarrow\widehat{{\cal H}}$.
The constant $\alpha$ is supposed to be very small, since the non-commutativity between $r$ and $\theta$ is not observed in everyday life. Therefore
$\alpha\ll 1$ which leads to
$\sin{\left(i\alpha\frac{\partial}{\partial r}\right)}\simeq
i\alpha\frac{\partial}{\partial r}$. After some algebraic
manipulations we find that $\widehat{{\cal H}}$ is given by

\begin{equation}
\widehat{{\cal H}}=4ki\alpha\left\{\left[1-\frac{\left(1-\frac{M}{r}\right)}{\left(1-\frac{2M}{r}\right)^{1/2}}\right]\frac{\partial}{\partial r}+\frac{M/2r^2}{\left(1-\frac{2M}{r}\right)^{3/2}}\right\}\,.
\end{equation}
We immediately see that this operator is anti-hermitian, therefore it also has real eigenvalues.
We suppose a eigenvector/eigenvalue equation as
$\widehat{{\cal H}}\psi=\epsilon\psi$ which leads to an equation of the form $\frac{\partial\psi}{\partial
r}+g(r)\psi=0$, where $g(r)$ is written as
$$g(r)=\left[1-\frac{\left(1-\frac{M}{r}\right)}{\left(1-\frac{2M}{r}\right)^{1/2}}\right]^{-1}\left[i\frac{\epsilon}{4k\alpha}+\frac{M/2r^2}{\left(1-\frac{2M}{r}\right)^{3/2}}\right]\,.$$
The quantity $\epsilon$ is the eigenvalue. We point out that in our unit system the Hamiltonian density is adimensional, as a consequence the energy has length dimension since it comes from a volume integration of the Hamiltonian density. Therefore the Hamiltonian eigenvalue is adimensional as well and then it is given by $\epsilon=E/M$, where $E$ is the observable of the field.

Since the above equation is a first order differential equation, its solution is

$$\psi=\psi_0\exp{\left(-\int g(r) dr\right)}\,,$$
where $\psi_0$ is a constant of integration. It can be chosen to normalize the solution. Let us analyze the consequences of such a solution in the limit $M<<r$. Then we find
$$\psi=\psi_0\exp{\left(-\frac{i\epsilon}{8k\alpha M}r^2\right)}\,,$$
in the next step we impose that the solution should assume the same values at the singularity points $r=0$ and $r=2M$, hence $\psi(0)=\psi(2M)$. It is well known that $E=M$ for Schwarzschild space-time which leads to the conclusion that $\epsilon=1$. We point out that the value of the gravitational energy yielded by TEGR is the classical observable, thus the eigenvalue of our quantum equation should fulfill such expectation. Therefore we finally have $$M=n\,m_0\,,$$ where $n$ is an integer once $k=1/16\pi$ and $m_0=\alpha/4$. Such a condition arises from the use of the boundary condition $\psi(0)=\psi(2M)$. In the international unit system we have $m_0=\frac{\alpha c^2}{4G}$ which is the quantum of matter.

\section{Conclusion}

In this article we have presented a formal procedure to construct a quantum theory of gravity. We have performed our calculations in the realm of Teleparallel gravity due to the arising of a proper energy-momentum vector as one of its fundamental features. We have used the Weyl quantization process to obtain operators out of classical quantities, then we establish an eigenvalue/eigenvector equation which reveals the quantum features of the field in the context of Schwarzschild space-time. Such quantum properties are obtained by the imposition of periodic conditions on the eigenfunction which is the solution of $\widehat{{\cal H}}\psi=\epsilon\psi$. This leads to the definition of $m_0$, the quantum of matter. Thus the black hole mass is quantized in terms of such a parameter. The quantum of matter, in the international unities, is written in terms of gravitational constant, speed of light and $\alpha$ which is a constant with dimension of length, introduced in the quantization process. Therefore, in order to give the order of magnitude of this new constant, we point out that every piece of matter is formed by electrons as its smallest mass constituents. We recall that others tiny constituents such as quarks are more massive than electrons. Hence we associate the quantum of matter to the electron's mass, this yields $\alpha\sim 10^{-56}\,m$. Bearing this in mind, we think that the electron may have different mechanisms to yield what it is observed, for instance, one responding for matter and another one for charge and spin. We also point out that the results obtained in this article were derived from a hamiltonian density, ${\cal H}=e\,t^{(0)0}$, which is not invariant under coordinate transformations. Such a feature is also present when one tries to quantize fields in a curved space-time. In fact it is not a problem at all, since we expect a break of the diffeomorphic symmetry when constructing a quantum theory of gravitation, once coordinates were changed to operators. This feature would lead to different equations for each coordinate system, however all of them should behave equally in the limit $M/r<<1$. Our results may be extended to fundamental particles since their line element can be described by Schwarzschild's solution in isotropic coordinates as obtained in reference~\cite{Katanaev:2012wc}.

%

%\bibliography{ref}
%\bibliographystyle{apsrev4-1}

\end{document}